\documentclass{ws-ijgmmp}
\usepackage{notoccite}
\usepackage[usenames,dvipsnames,svgnames]{xcolor}
\usepackage[colorlinks=true,
      linkcolor=red,
      urlcolor=gray,
      citecolor=blue]{hyperref}
\usepackage[]{cite}
\usepackage{float}

\begin{document}

\markboth{Orlando Luongo, Stefano Mancini}
{Entanglement in model independent cosmological scenario}

\catchline{}{}{}{}{}

\title{Entanglement in model independent cosmological scenario}

\author{Orlando Luongo}
\address{Istituto Nazionale di Fisica Nucleare (INFN), Laboratori Nazionali di Frascati, Via E. Fermi 40, 00044 Frascati, Italy.\\
School of Science and Technology, University of Camerino, Via Madonna delle Carceri 9, 62032 Camerino, Italy.\\
NNLOT, Al-Farabi Kazakh National University, Al-Farabi av. 71, 050040 Almaty, Kazakhstan.}

\author{Stefano Mancini}
\address{School of Science and Technology, University of Camerino, Via Madonna delle Carceri 9, 62032 Camerino, Italy.\\
Istituto Nazionale di Fisica Nucleare (INFN), Sezione Perugia, Via A. Pascoli, 06123 Perugia, Italy.}

%%%%%%%%%%%%%%%%%%%%%%%%%%%%%%%%%%%%%%%%%%%%%%%%%%%%%%%%%%%%%%%%%%%%%
%%%%%%%%%%%%%%%%%%%%%%%%%%%%%%%%%%%%%%%%%%%%%%%%%%%%%%%%%%%%%%%%%%%%%
%%%%%%%%%%%%%%%%%%%%%%%%%%%%%%%%%%%%%%%%%%%%%%%%%%%%%%%%%%%%%%%%%%%%%

\maketitle

\begin{abstract}
We propose a \emph{model-independent} approach to study entanglement creation due to the dynamics between two asymptotic quantum regimes in the framework of homogeneous and isotropic universe.
We realize it by Pad\'e expansion to reconstruct the functional form of scale factor,
rather than postulating it a priori.This amounts to fix the Pad\'e approximants constraining the free parameters in terms of current cosmic observations.
Assuming fermions, we solve the Dirac equation for massive particles and we investigate entanglement entropy in terms of modes $k$ and mass $m$. We consider two rational approximations of (1,1) and (1,2) orders, which turn out to be the most suitable choices for guaranteeing cosmic bounds.
Our results show qualitative agreement with those known in literature
and arising from toy models, but with sensible
quantitative discrepancies. Moreover, our outcomes are model independent reconstructions which show that any higher orders departing from (1,1) do not  significantly modify particle-antiparticle production (hence entanglement), if cosmic bounds are taken into account.
\end{abstract}

%%%%%%%%%%%%%%%%%%%%%%%%%%%%%%%%%%%%%%%%%%%%%%%%%%%%%%%%%%%%%%%%%%%%%
%%%%%%%%%%%%%%%%%%%%%%%%%%%%%%%%%%%%%%%%%%%%%%%%%%%%%%%%%%%%%%%%%%%%%
%%%%%%%%%%%%%%%%%%%%%%%%%%%%%%%%%%%%%%%%%%%%%%%%%%%%%%%%%%%%%%%%%%%%%

%\pacs{03.65.Ud, 04.62.+v, 98.80.-k}


%%%%%%%%%%%%%%%%%%%%%%%%%%%%%%%%%%%%%%%%%%%%%%%%%%%%%%%%%%%%%%%%%%%%%
%%%%%%%%%%%%%%%%%%%%%%%%%%%%%%%%%%%%%%%%%%%%%%%%%%%%%%%%%%%%%%%%%%%%%
%%%%%%%%%%%%%%%%%%%%%%%%%%%%%%%%%%%%%%%%%%%%%%%%%%%%%%%%%%%%%%%%%%%%%

\section{Introduction}

Quantum correlations, also known as \emph{entanglement}, captured renewed attention after the development of quantum theory of information \cite{entareview}. Recently it has been realized
that entanglement also arises in cosmological scenarios \cite{MM12}.
It is actually related to the mechanism of particle-antiparticle
production during cosmic evolution, a phenomenon pointed out some time ago \cite{Parker}.
To ascertain that, given the difficulties in solving the dynamics, especially for what
concern the Dirac field, it is often resorted to
cosmological toy models.

Widely employed models are due to Duncan \cite{D78} and Birrell and Davies \cite{BD80}, where
the universe is viewed as a dynamical system obeying the hypothesis of homogeneity and isotropy, i.e. satisfying the Friedmann-Robertson-Walker (FRW)
line element $ds^2= d\tau^2-a(\tau)^2\left[dr^2 + r^2( d\theta^2+ \sin^2 \theta\ d\phi^2)\right]$. The scale factor $a(t)$ was devised to simply show an accelerated expansion phase of the universe
between two stationary asymptotic regions, corresponding to the early and the late universe.

Nowadays the most accredited paradigm is the cosmological concordance model, named $\Lambda$CDM
 \cite{copeland}.
There, the component responsible for the cosmic speed up is the cosmological constant $\Lambda$ and manifests a negative equation of state \cite{rev1bis,rev2,dudu} providing gravitational repulsive effects\footnote{Alternatives to the concordance model are commonly named \emph{dark energy} scenarios, in which the equation of state is a varying function. } \cite{ioequew}.
Thus in order to study entanglement there is a need to elaborate a \emph{model-independent} approach that  formulate the scale factor evolution within the concordance paradigm, since the term pushing the universe up is still undetermined. Any model-independent treatment needs to leave unset the free parameters that can be bounded  from experimental observations to agree with both late and early dynamics \cite{io}.

In this work, we propose the Pad\'e rational expansions \cite{padoriginal1,padoriginal2} as \emph{model-independent} reconstruction for the scale factor $a(t)$ in order to analyze the entanglement production. The use of Pad\'e approximations is motivated by the fact that they represent the most suitable way to build up a high order convergent series on $a(t)$ which converges at both late and early stages \cite{altropaddy1,altropaddy2,altropaddy3,altropaddy4,altropaddy5,ulteriore}. We show that  the involved polynomials might be rational. They well adapt to describe the universe without assuming the functional form of the Hubble parameter at arbitrary times with two peculiar expansion orders, namely (1,1) and (1,2). We fix the free parameters in terms of modern observations. We thus use the Pad\'e polynomials to study the entanglement production.

Actually we consider quantum states associated to matter at two epochs: the first concerning earliest phases, i.e. before expansion, whereas the second as expansion is over, i.e. at the future.
Then we introduce the dynamical process in a way that
the functional form of $a(t)$ is reconstructed from recent kinematic data coming from cosmography, without postulating the model \emph{a priori}.
We solve the Dirac equation for massive particles and we compute the Bogolubov transformations between the two involved epochs. We  evaluate the particle-antiparticle entanglement as subsystem (particle) entropy in terms of mode $k$ and mass $m$.

Particularly, models of Refs. \cite{D78,BD80} are featured by introducing a \emph{ad hoc} scale factors which do not match any cosmological requirements. The models depend upon free constants, which have been introduced to stress the expansion rate and the volume variation. The Pad\'e expansions do not fix \emph{a priori} the number of free parameters. Although the set of free parameters may be in principle larger, each term simply corresponds to a series coefficient and is not imposed by hand.

In particular, we recover the Duncan's model \cite{D78} from (1,1) Pad\'e approximant. Further,
the entanglement obtainable from it (see e.g. \cite{Fetal10})  is in good agreement with current cosmic observations only for small and high values of $k$. For intermediate and large values of $k$, higher order Pad\'e approximants do not significantly modify the entanglement production got from (1,1) approximation.

The paper is structured as follows. In Sec. II, we present our model-independent approach based on
Pad\'e approximations for the universe dynamics. Actually we show how to construct a viable approximation to $a(t)$ motivating our choice through the use of recent developments in the literature. In Sec. III, we describe the simplest solution to $a(t)$ as function of Pad\'e polynomials. Using it we derive analytical solutions of Dirac equation in terms of Hypergeometric functions of second kinds.
In turn this allows us to compute the entanglement production.
In Sec. IV we consider higher order solution to $a(t)$ as function of Pad\'e polynomials and compute numerically the entanglement production.
We give also a detailed interpretation of the so-obtained results. Finally, we report conclusions and perspectives in Sec. V.

%%%%%%%%%%%%%%%%%%%%%%%%%%%%%%%%%%%%%%%%%%%%%%%%%%%%%%%%%%%%%%%%%%%%%

\section{The Pad\'e approximation as model-independent technique}\label{quantummodel}

The simplest approach in defining rational approximations to the scale factor is offered by the Pad\'e series \cite{padoriginal1,padoriginal2}. The basic demand over the Pad\'e construction is to overcome the model-dependence issue on $a(\tau)$ as one postulates the functional time-dependence of
$a(\tau)$. Indeed, expanding in Taylor series around a precise time, say $\tau_*$, leads to powers of $\tau$ which are defined only around $\tau\approx \tau_*$.
In other words, the $a(\tau)$ Taylor series, taking $\tau_*$ as reference time, gives:
\begin{equation} \label{aint}
    a(\tau) \equiv 1+\sum_{k=1}^{\infty}\frac{1}{k !}\frac{d^k a}{dt^k}\Big|_{\tau=\tau_*}(\tau-\tau_*)^k \,,
\end{equation}
where the coefficients are named \emph{cosmographic parameters}.
These coefficients are well defined as they are fixed at $\tau=\tau_*$
and give information on the universe dynamics at large scales.\footnote{In observational cosmology, it is customary to take $\tau_*$ as our time.}  Indeed, the advantage of expanding $a(\tau)$ is that the cosmographic parameters may be directly measured with cosmic data, i.e. without fixing \emph{a priori} the cosmological model. Approaches toward the definition of the cosmographic series and bounds over it have been extensively investigated in the literature \cite{cosmografia,cosmografia2}.

From a pure experimental point of view, this procedure if compared directly with data, introduces errors into the analysis since the employed formulae only represent approximations to the true expressions. Moderating the caveat may involve higher orders of the Taylor series, but this comes at the expense of introducing more fitting parameters. Furthermore the flat-to-flat behavior is not preserved. This would considerably complicate the corresponding statistical analysis \cite{avilesorl}.

To go beyond the expected convergence range of Taylor series\footnote{This problem is known
in the literature as convergence problem \cite{si1}.}, one can shift to the Pad\'e series.
Let us recall that for a generic function $f(z)$ the Pad\'e approximant with fixed orders $(m,n)$ is defined as:
\begin{equation} \label{facy}
  P_{mn}(z) = \frac{a_0 + a_1 z + a_2 z^2 + \ldots + a_m z^m}{1+b_1 z + b_2 z^2
    + \ldots + b_n z^n} \,.
\end{equation}
A more general $(M,N)$  Pad{\'e} approximant may be written under the form:
\begin{equation}
P_{M, N}(z)=\dfrac{\sum_{m=0}^{M}a_m Z^m}{1+\sum_{n=1}^{N}b_n Z^n}\,,
\label{eq:def_Pade}
\end{equation}
where $Z=Z(z)$ is an arbitrary function of the independent variable (for example the redshift $z$).

The aforementioned definition of Pad\'e series is here used to match cosmic rulers, assuming that Pad\'e expansions are equivalent to the standard Taylor series up to the highest possible orders. To do so, we require
\begin{subequations}
\begin{align}
&P_{M,N}(0)=f(0)\ , \\
&P_{M,N}'(0)=f'(0)\ ,\\
&\vdots	\nonumber\\	
&P_{M,N}^{(M+N)}(0)=f^{(M+N)}(0)\ .
\end{align}
\end{subequations}
In principle the net number of total unknown terms is $M+N+1$, defined as the sum between the $M+1$ coefficients of the numerator and $N$ terms of the denominator. In turn, we have:
\begin{equation}
\sum_{i=0}^\infty c_iz^i=\dfrac{\sum_{m=0}^{M}a_m z^m}{1+\sum_{n=1}^{N}b_n z^n}+\mathcal{O}(z^{M+N+1})
\end{equation}
and then
\begin{align}
(1+b_1z+\hdots +b_Nz^N)(c_0+c_1z+\hdots)=
a_0+a_1z+\hdots+a_Mz^M +\mathcal{O}(z^{M+N+1})\,.
\label{coeff}
\end{align}
Plugging together the terms with the same power, one gets a set of $M+N+1$ equations for any $M+N+1$ unknown coefficients $a_i$ and $b_i$. For $z \gg 1$ the series converges and thus it candidates to overcome Taylor' series divergencies.

Following the standard recipe of Pad\'e polynomials the most suitable set of exponents requires that:

\begin{enumerate}
  \item Pad\'e series behaves smoothly with high redshift cosmic data;
  \item Pad\'e series minimize systematics in approximating the universe expansion history;
  \item orders $M,N$ might be comparable to converge to constants asymptotically;
  \item convergence is calibrated \emph{a posteriori} by means of data surveys;
  \item coefficients are fixed to avoid possible poles;
  \item Pad\'e series might agree with previous approaches\footnote{This  gives a robust physical explanation to the toy model discussed in \cite{D78}.}.
\end{enumerate}

We are interested in approximating \emph{directly} the scale factor $a(\tau)$ using the Pad\'e series, in agreement with current cosmological bounds. Thus, the simplest orders that seem to be viable for approximating $a(\tau)$ are $(1,1)$ and $(1,2)$ Pad\'e series respectively. These two orders well adapt to the six conditions above described.

In our scenario, the Pad\'e approximation will be built up in terms of a positive-definite functional
$t(\tau)=-\ln\tau$. The functional dependence is chosen to match time $t\in(-\infty,+\infty)$ with time $\tau\in[0,+\infty)$, so to respect the fact that $\tau_0=0$ is current time, while
$\tau=+\infty$ is the remote past. Moreover, one can notice that $\tau_*$ corresponds to the cosmographic time at which the scale factor is \emph{conventionally} normalized to $a(\tau_*)\neq1$. In addition, it is also demanded the stationarity of the spacetime for $t\to\pm\infty$
as in Refs. \cite{D78} and \cite{BD80}. This enables to have two natural quantizations of the field associated with two Fock spaces \cite{bah}.

%%%%%%%%%%%%%%%%%%%%%%%%%%%%%%%%%%%%%%%%%%%

\section{Entanglement from the scale factor $(1,1)$ Pad\'e approximant}
\label{sec:pade11}

In the previous section, we introduced the main reasons behind the choice of Pad\'e expansions instead of using the standard Taylor approach. We highlight that the smallest orders of Pad\'e polynomials, namely the $(1,1)$ and $(1,2)$, as applied to luminosity distance $d_L$ significatively improve the convergence radius.
Hence, we can start adopting the expansion
\begin{equation}\label{expansionnuova}
a_{(1,1)}(\tau)=\frac{\beta_0+\beta_1 \tau}{1+\beta_2 \tau}\,.
\end{equation}
In terms of $t$ we have
\begin{subequations}\label{condix}
\begin{align}
a_{(1,1)}(t=-\infty)=\frac{\beta_1}{\beta_2}\,,\\
a_{(1,1)}(t=+\infty)=\beta_0\,,
\end{align}
\end{subequations}
as well as $da_{(1,1)}/dt|_{t=-\infty}=da_{(1,1)}/dt|_{t=+\infty}=0$.

Physically speaking one can assume the Big Bang time, as the one at which quantum effects are remarkable. However, the choice of this value does not influence our approach. Indeed, it only gives a shift to the weight and strength of all curves and functions that we are working with, leaving unaltered the physical properties behind the choice of our Pad\'e formalism. As a matter of fact, we note that Eqs. \eqref{condix} represent conditions which work well only in certain epochs of universe's evolution. They cannot be used for the whole large scale dynamics.

Below we shall consider matter field $\psi$ (of mass $m$ and spin $\frac{1}{2}$) in 1+1 spacetime. In the far future we assume a Dirac field, with spins $\frac{1}{2}$ associated to each particle. A simple and immediate request is the existence of a Hilbert space both in the past flat region ($t=-\infty$) and in the future flat epoch ($t=+\infty$), with a suitable choice concerning basis vectors for each era.

%%%%%%%%%%%%%%%%%%%%%%%%%%%%%%%%%%%%%%%%%%%%%%%%%%%

\subsection{Limits over (1,1) Pad\'e expansion of $a(t)$}

We need to fix the parameters in \eqref{expansionnuova} in the most suitable way, i.e. to agree with observational properties associated to scale factor evolution at late and early times. In particular, computing the kinematics of our model, one immediately requires a de-Sitter-like phase for $t=0$, to account for the large scale acceleration. Indeed, requiring the definition of Hubble rate ($H$) and deceleration parameter ($q$), respectively given by:
\begin{eqnarray}\label{HQ}
H\equiv\frac{\dot a}{a}\qquad q\equiv-1-\frac{\dot H}{H^2}\,,
\end{eqnarray}
one immediately finds that $H$ and $q$ read:
\begin{subequations}\label{h0andq0}
\begin{align}
H_0&=\frac{\beta_0}{\beta_0 + \beta_1} - \frac{1}{1 + \beta_2}\,,\label{h0andq01}\\
q_0&=-\frac{(\beta_0 + \beta_1)(\beta_2-1)}{\beta_0 \beta_2-\beta_1}\,,\label{h0andq02}
\end{align}
\end{subequations}
where the subscript $0$ refers to $H$ and $q$ as computed at current time. By virtue of Eqs. \eqref{h0andq0}, one can remove some arbitrariness on the constants, requiring that
\begin{eqnarray}\label{limitsnc}
\beta_0&>0\,,\\
\beta_1&\neq\beta_2\,.
\end{eqnarray}
The first condition, i.e. $\beta_0>0$, states that the universe's volume increases as byproduct of the standard Big Bang model. The second request over $\beta_1$ and $\beta_2$ is less stringent and represents our arbitrariness to frame how  cosmic expansion rate behaves far from asymptotic regimes. In fact, asymptotically it is licit to relax the second condition, having a non-accelerating (but expanding) universe in the limits $\tau\rightarrow0$ and $\tau\rightarrow\infty$, while it is not licit to take $\beta_0<1$ even asymptotically. Bearing in mind these bounds, without losing generality we require $\beta_1$ and $\beta_2$ to be positive-definite in general and $\beta_0>1$. Moreover to guaranteeing the physical robustness of Eqs. \eqref{h0andq0} we take the range of values satisfying the condition
\begin{equation}\label{betatutto}
\beta_0\beta_2>\beta_1\,.
\end{equation}
The request \eqref{betatutto} over $\beta_0, \beta_1$ and $\beta_2$ implies a degeneracy among the coefficients \cite{degenerazione1,degenerazione2}. Thus a mathematical trick useful to characterize the
scale factor evolution at large scales can be based on recasting the constants in Eq. \eqref{expansionnuova}. In particular, if we consider
\begin{equation}\label{hhh}
\beta_2e^{-t}\rightarrow e^{-T}\,,
\end{equation}
we have $t=T+\ln\beta_2$ and so $a_{1,1}(t)$ transforms as
\begin{equation}\label{massatrasformata}
a_{1,1}(T)=\frac{\beta_0+\beta^\prime e^{-T} }{1 + e^{-T} }\,.
\end{equation}
This choice of time variable disguises $\beta_1$ and $\beta_2$ through a single constant, namely
$\beta^\prime\equiv\beta_1/\beta_2$. It only represents a time-shift and does not influence the age of the universe. In fact, the latter is given by a precise value of $\tau$ in the interval $\in[0,+\infty)$, which remains unchanged. Notice that the derivatives with respect to $t$ are equivalent to the derivatives with respect to $T$. Furthermore, the asymptotic limits on $t$ are the same on $T$, i.e. as $t\rightarrow\infty, T\rightarrow\infty$. Hence below we can interchange $T$ with $t$ without affecting the final outcomes.

\subsection{Getting solutions for Dirac's field}

On curved space-time, the Dirac equation for the field $\psi$ of mass $m$ reads:
\begin{align}\label{dirac}
\Big[\tilde\gamma^{\mu}D_{\mu}\,+\,m\Big]\psi=0\,,
\end{align}
where, working on a FRW space-time, we defined $\tilde\gamma^{\mu}\equiv[a(t)]^{-1}\gamma^\mu$ that are the
re-scaled $2\times 2$ spinor matrices and $D_{\mu}$ the covariant derivative.
Employing the auxiliary field $\varphi$, we may look for solutions under the form \cite{birrell}:
\begin{align}
  \psi &= a^{-1/2}(\gamma^{\nu}\partial_{\nu}-M)\varphi\,,
\end{align}
with effective mass given by: $M = ma(t)$. So that, recasting the Dirac equation \eqref{dirac} by
\begin{equation}\label{6788}
\Box\varphi-\gamma^{0}\dot{M}\varphi
-M^{2}\varphi=0\,,
\end{equation}
where $\Box\equiv g^{\mu\nu}\partial_{\mu}\partial_{\nu}$, we get the corresponding solutions as:
\begin{eqnarray}
&  \varphi^{(-)}\equiv N^{(-)}f^{(-)}(t) u e^{ i k x}, \label{varphi1}\\
&  \varphi^{(+)}\equiv N^{(+)}f^{(+)}(t) v e^{ i k x}, \label{varphi2}
\end{eqnarray}
with $u$, $v$ the flat spinors and $k$ the momentum.
After some algebra, one can get the differential equation for the time dependent
functions $f^{(\pm)}$:
\begin{equation}
\ddot{f}^{(\pm)}+\left(  k^{2}+M^{2}\pm i\dot{M}\right)f^{(\pm)}=0\,. \label{feq1000}
\end{equation}
Let us now define $f^{(\pm)}_{in/out}$ and ${f^{(\pm)}}^*_{in/out}$ the solutions behaving
as positive and negative frequency modes with respect to time~$t$
near the asymptotic past/future, i.e.
\begin{equation}\label{feq}
\dot{f}^{(\pm)}_{in/out}(t) \approx -i E_{in/out} f^{(\pm)}_{in/out}(t),
\end{equation}
where
\begin{subequations}\label{EM}
\begin{align}
E_{in/out}&\equiv\sqrt{k^2+M_{in/out}^2},\\
M_{in/out}&\equiv m a(t \to -\infty/+\infty).
\end{align}
\end{subequations}
Then, from Eq. \eqref{feq}, we have
\begin{subequations}\label{asymptsol}
\begin{align}
f_{in}^{(\pm)}(t)&=e^{i\delta t}\,\,{}_2\mathcal{F}_1\Big[\theta_1^\pm,\theta_2^\mp,\theta_3^{-}, \ell_1(t)\Big],\\
f_{out}^{(\pm)}(t)&=e^{i\delta t}\,\,{}_2\mathcal{F}_1\Big[\theta_1^\pm,\theta_2^\mp,\theta_3^{+}, \ell_2(t)\Big].
\end{align}
\end{subequations}
where ${}_2{\cal F}_{1}$ denotes the Hypergeometric functions of the second kind.
The coefficients $\theta_1^\pm, \theta_2^\pm$ and $\theta_3^\pm$ are defined as
\begin{subequations}
\begin{align}
\theta_1^\pm&\equiv1+i\Big[2E_{-}\pm m(\beta_0-\beta^\prime)\Big]\,,\\
\theta_2^\pm&\equiv\Big[2E_{-}\pm m(\beta_0-\beta^\prime)\Big]\,,\\
\theta_3^\pm&\equiv1\pm2iE_{out/in}\,,
\end{align}
\end{subequations}
with
\begin{equation}
E_{\pm}\equiv\frac{1}{2}\left(E_{out}\pm E_{in}\right).
\end{equation}
Furthermore it is $\ell_2\equiv1-\ell_1$ with
\begin{equation}
\ell_1\equiv\frac{\exp\left[{t\over2}\right]}{\exp\left[{t\over2}\right]+\exp\left[-{t\over2}\right]}\,.
\end{equation}
Going back to Eq. \eqref{asymptsol}, the function $\delta$ depends in principle upon the whole set of parameters defined in Eqs. \eqref{expansionnuova} and \eqref{EM}, i.e. $\delta\equiv \delta(E_{out},E_{in},\beta_0,\beta^\prime)$. Cumbersome algebra shows it does not depend on $\beta_0$ and $\beta^\prime$ since it turns out to be a phase factor. Indeed it reads
\begin{equation}
\delta\equiv E_{+}t+2E_{-}\ln\left(2\cosh (t/2)\right).
\end{equation}
Notice that in the solutions \eqref{asymptsol} of Dirac equation the Hypergeometric functions
${}_2{\cal F}_{1}$ describe the slowly varying (on time) part
 of the mode functions, contrasted to the oscillatory behavior of the unimodular factor $e^{i\delta t}$.

%%%%%%%%%%%%%%%%%%%%%%%%%%%%%%%%%%%%%%%%%%%%%%%%%%%

\subsection{Particle-antiparticle production}

As one expands the field $\psi$ over spinors in \emph{input} and \emph{output} regions, it is possible to relate the coefficients of such expansions,
namely the in-out ladder operators for particles and
antiparticles (denoted by $a$, $a^\dag$ and $b$, $b^\dag$ respectively),
by Bogolubov transformations \cite{D78}:
\begin{subequations}\label{ab}
\begin{align}
a_{out}(k)&=\alpha(k)a_{in}(k)-\beta(k)b_{in}^\dag(-k),\\
b_{out}^\dag(-k)&=\beta^*(k)a_{in}(k)+\alpha^*(k)b_{in}^\dag(-k),
\end{align}
\end{subequations}
with coefficients satisfying $|\alpha|^2+|\beta|^2=1$ and $\alpha\beta^*-\alpha^*\beta=0$.

Analogously, Bogolubov transformations interconnect the solutions $f^{(\pm)}_{in/out}$ giving
\begin{equation}\label{fbog}
f^{(\pm)}_{in}(t)=A^{(\pm)}(k) f^{(\pm)}_{out}(t)+ B^{(\pm)}(k) f^{(\mp)*}_{out}(t)\,.
\end{equation}
Clearly the coefficients $A^{(\pm)},B^{(\pm)}$ are related to $\alpha,\beta$, in particular it results \cite{MPM14}
\begin{equation}\label{alphaA}
|\alpha(k)|^2=\frac{E_{out}}{E_{in}}\left(\frac{E_{in}-M_{in}}{E_{out}-M_{out}}\right)\Big| A^{(-)}(k)
\Big|^2\,.
\end{equation}
Additionally, the Bogolubov coefficients can be related to particle-antiparticle production. In fact letting $n$ be the density of particles per mode at the output, we have
\begin{align}\label{betaN}
n(k)&\equiv{}_{in}\langle 0_k,0_{-k}| \left(a_{out}^\dag(k)a_{out}(k)+b^\dag_{out}(-k)b_{out}(-k)\right)
|0_k,0_{-k}\rangle_{in}\notag\\
&=2|\beta(k)|^2 =2\left(1- |\alpha(k)|^2\right),
\end{align}
where $|0_k,0_{-k}\rangle_{in}$ denotes the input vacuum (the subscript $k$, resp $-k$, refers to the particle, resp. antiparticle).
Notice that $0\le n \le 2$ with the maximum accounting for the presence of both particle and antiparticle.

We rewrite Eq. \eqref{fbog} by
\begin{eqnarray}
{}_2\mathcal{F}_1\left(\theta_1^{-}, \theta_2^{+},
\theta_3^{-}, \ell_1\right)=A^{(-)}
{}_2\mathcal{F}_1\left(\theta_1^{-}, \theta_2^{+},
\theta_3^{+}, \ell_2\right)+B^{(-)} e^{2i\delta}
{}_2\mathcal{F}_1\left(\theta_1^{+}, \theta_2^{-},
\theta_3^{+}, \ell_2\right)^*,\notag\\
\end{eqnarray}
and furthermore we have
\begin{align}
{}_2\mathcal{F}_1(\theta_1^{-},\theta_2^{+},\theta_3^{-},\ell_1)&=A^{(-)} {}_2\mathcal{F}_1(\theta_1^{-},\theta_2^{+},\theta_1^{-}+\theta_2^{+}-\theta_3^{-}+1,\ell_2)\notag\\
&\hspace{-1cm}
+B^{(-)}\, \ell_2^{\theta_3^{-}-\theta_1^{-}-\theta_2^{+}}\,\ell_1^{1-\theta_3^{-} }\,{}_2\mathcal{F}_1(1+\theta_2^{+},\theta_1^{-}-1,\theta_1^{-}+\theta_2^{+}-\theta_3^{-}+1,\ell_2)^*\,.\notag\\
\end{align}
Hence, given that $\ell_1\in[0,1]$, we can use the property of hypergeometric functions of second kind
\bigskip
\begin{align}\label{uno5}
{}_2\mathcal{F}_1(\theta_1^-,\theta_2^+,\theta_3^-,\ell_1)&=
A^{(-)} {}_2\mathcal{F}_1(\theta_1^-,\theta_2^+,\theta_1^-+\theta_2^+-\theta_3^-+1,\ell_2)\notag\\
&\hspace{-2cm}+B^{(-)}\, \ell_2^{\theta_3^--\theta_1^--\theta_2^+}\, \ell_1^{1-\theta_3^- }\,
{}_2\mathcal{F}_1\Big[1+\theta_2^{^+,*},\theta_1^{^-,*}-1,\theta_1^{^-,*}+\theta_2^{^+,*}-\theta_3^{^-,*}+1, \ell_2(t)\Big]\,,\notag\\
\end{align}
which is compatible with Eq. \eqref{wittm} and comes from the fact that
${}_2\mathcal{F}_1(\theta_1^-,\theta_2^+,\theta_3^-,\ell_1)^*={}_2\mathcal{F}_1(\theta_1^{^-,*},\theta_2^{^+,*},\theta_3^{^-,*},\ell_1)$ in the interval $\ell_1\in[0,1]$.

Adopting the symmetry of ${}_2\mathcal{F}_1$
with respect to the exchange of the first two arguments we get
\begin{align}\label{uno6}
{}_2\mathcal{F}_1\left(\theta_1^{-}, \theta_2^{+},
\theta_3^{-}, \ell_1\right)&=A^{(-)} {}_2\mathcal{F}_1(\theta_1^-,\theta_2^+,\theta_1^-+\theta_2^+-\theta_3^-+1,\ell_2)\notag\\
&\hspace{-1cm}+B^{(-)}\, \ell_2^{\theta_3^--\theta_1^--\theta_2^+}\, \ell_1^{1-\theta_3^- }\,
{}_2\mathcal{F}_1\Big[1-\theta_1^-,1-\theta_2^+,\theta_3^--\theta_1^--\theta_2^++1,\ell_2\Big].
\notag\\
\end{align}
Moreover since
\begin{align}
&\ell_1^{1-\theta_3^- }{}_2\mathcal{F}_1
(1-\theta_1^-,1-\theta_2^+,\theta_3^--\theta_1^--\theta_2^++1,\ell_2)\notag\\
&\hspace{2cm}={}_2\mathcal{F}_1(\theta_3^--\theta_1^-,\theta_3^--\theta_2^+,\theta_3^--\theta_1^--\theta_2^++1,\ell_2),
\end{align}
it is licit to write
\bigskip
\begin{align}\label{uno7}
{}_2\mathcal{F}_1\left(\theta_1^{-}, \theta_2^{+},\theta_3^{-}, \ell_1\right)&=A^{(-)}
{}_2{\cal F}_1
(\theta_1^-,\theta_2^+,\theta_1^-+\theta_2^+-\theta_3^-+1,\ell_2)\notag\\
&\hspace{-1cm}+B^{(-)}\, \ell_2^{\theta_3^--\theta_1^--\theta_2^+} \,
{}_2\mathcal{F}_1\Big[\theta_3^--\theta_1^-,\theta_3^--\theta_2^+,\theta_3^--\theta_1^--\theta_2^++1,\ell_2\Big].\notag\\
\end{align}
We need $A^{(-)}$ to compute $n(k)$ as shown in Eqs.\eqref{alphaA}-\eqref{betaN}. Hence,
using Eq.\eqref{uno7} and the property \eqref{A1} of Hypergeometric function reported in Appendix A, we get
\begin{align}\label{ameno}
A^{(-)}(k) =\frac{\Gamma(1-2iE_{in})\Gamma
(-2iE_{out})}{\Gamma(1-i2E_{+}-im(\beta_0-\beta^\prime))
\Gamma(-2iE_{+}+ im(\beta_0-\beta^\prime))},
\end{align}
and
\begin{align}\label{bimeno}
B^{(-)}(k) =\frac{\Gamma(1-2iE_{in})\Gamma(2iE_{out})}{\Gamma(1+2iE_{-}-im(\beta_0-\beta^\prime))
\Gamma(2iE_{-}+im(\beta_0-\beta^\prime))}.
\end{align}
Finally, using \eqref{alphaA} and \eqref{betaN}, we arrive at
\begin{align}
n(k)=2\Bigg[1-&\frac{E_{out}}{E_{in}}\left(
\frac{E_{in}-M_{in}}{E_{out}-M_{out}}\right)\notag\\
&\times
\left|
\frac{\Gamma(1-2iE_{in})\Gamma
(-2iE_{out})}{\Gamma(1-i2E_{+}-im(\beta_0-\beta^\prime))
\Gamma(-2iE_{+}+ im(\beta_0-\beta^\prime))}
\right|^2\Bigg].
\end{align}

%%%%%%%%%%%%%%%%%%%%%%%%%%%%%%%%%%%%%%%%%%%%%%%%%%%%%%%

\subsection{Entanglement entropy}

By reverting Eqs.\eqref{ab}, we have
\begin{equation}\label{abrev}
a_{in}(k)=\alpha^*(k) a_{out}(k)+\beta(k) b^\dag_{out}(k),
\end{equation}
Furthermore, by means of the Schmidt decomposition (see e.g. \cite{entareview}), the input vacuum can be rewritten as
\begin{equation}
|0_k,0_{-k}\rangle_{in}=c_0 |0_k,0_{-k}\rangle_{out}+ c_1 |1_k,1_{-k}\rangle_{out}.
\end{equation}
Now $a_{in}(k) |0_k,0_{-k}\rangle_{in}=0$ implies, by means of \eqref{abrev},
\begin{equation}
\left(\beta c_0 + \alpha^* c_1\right) |0_k,1_{-k}\rangle_{out}=0,
\end{equation}
that is
\begin{equation}
c_1=-\frac{\beta}{\alpha^*}c_0.
\end{equation}
Also, by normalization condition, it is $c_0=\sqrt{1-|\beta|^2}$.
As consequence, for the mode $k$, the input vacuum is seen at the output as the following particle-antiparticle state
\begin{equation}\label{papstate}
|\Psi\rangle_{out}=\sqrt{1-|\beta|^2}\left( |0_k,0_{-k}\rangle_{out}-\frac{\beta}{\alpha^*} |1_k,1_{-k}\rangle_{out}
\right).
\end{equation}
Its reduced, e.g. particle, state reads
\begin{subequations}\label{subsys}
\begin{align}
\rho_{out,\,k}&={\rm Tr}_{-k}\left(|\Psi\rangle_{out}\langle\Psi|\right)\\
&=\left(1-|\beta|^2\right)\left( |0_k\rangle_{out}\langle 0_k|-\left|\frac{\beta}{\alpha^*}\right|^2
|1_k\rangle_{out}\langle 1_k|
\right)\\
&=\left(1-\frac{n}{2}\right) |0_k\rangle_{out}\langle 0_k|+\frac{n}{2}
|1_k\rangle_{out}\langle 1_k|,
\end{align}
\end{subequations}
where in the last line we have used \eqref{betaN}.

\begin{figure}[h!]
\begin{center}
\includegraphics[width=3.2in]{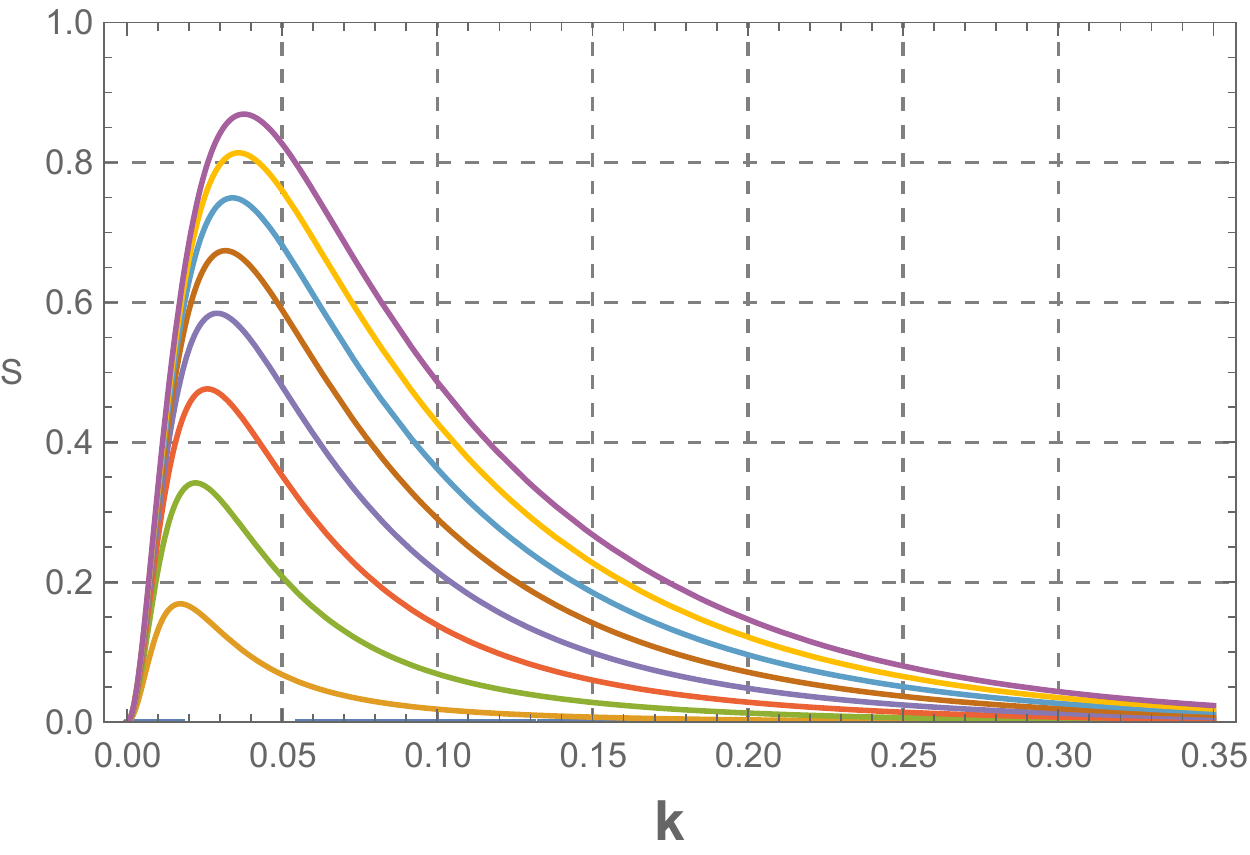}
\includegraphics[width=3.2in]{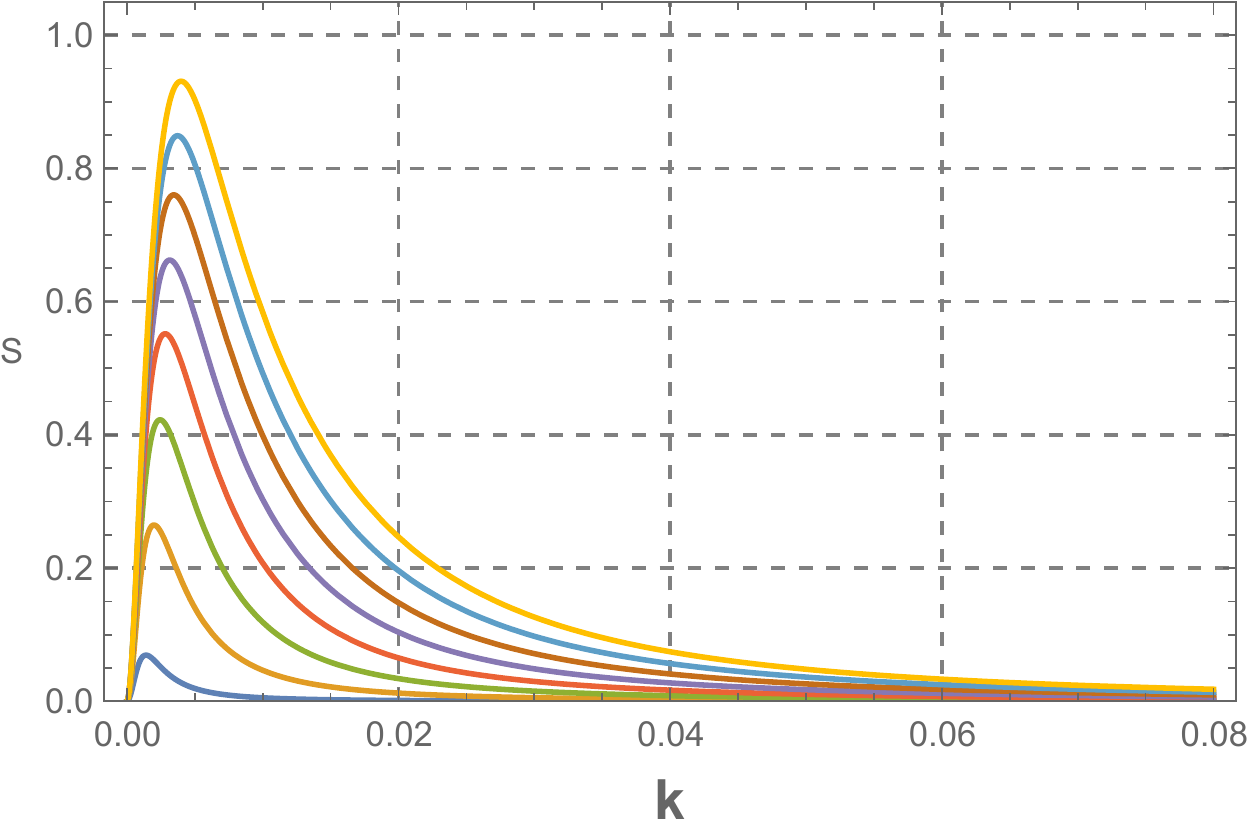}
\caption{Behavior of $S$ vs momentum $k$. Here, we consider $m=0.01$ (left) and $m=0.001$ (right). Curves from top to bottom correspond to values: $\beta_0\in[1,10]$
with step 1 and $\beta^\prime=1$.}
\label{esse}
\end{center}
\end{figure}

Being \eqref{papstate} a pure (bipartite) state a reliable measure of its entanglement is the von Neumann entropy
of the subsystem state \eqref{subsys} (see e.g. \cite{MPM14,PMM16}). This reads
\begin{equation}
S_{out}=-\frac{n}{2}\log_2\frac{n}{2}-\left(1-\frac{n}{2}\right)\log_2\left(1-\frac{n}{2}\right).
\end{equation}
The behavior of $S$ in terms of $k$ is reported in Figs. \eqref{esse}.

We can see that the entropy, as the momentum increases from zero on, reaches a maximum
and then decreases. Since we are interested in asymptotic regimes, we take $\beta^\prime=1$ and note the curves are broadened (and with higher maxima) as far as
the free parameter of the scale factor fulfills the condition: $\beta_0>1$.

Additionally if the mass increases the entropy production gradually tends to be suppressed. This behavior is depending also on $\beta_0$, as above discussed and degenerates with $m$. In fact, as $m$ tends to be smaller, the entropy is suppressed away at smaller $k$.

%%%%%%%%%%%%%%%%%%%%%%%%%%%%%%%%%%%%%%%%%%%%%%%%%%%%%%%%%%%%%%%%%%%%%%%%%%%%%%%%%%%%%%%%

\section{Entanglement from the scale factor $(1,2)$ Pad\'e approximant}
\label{sec:Pade12}

By referring to Eq. \eqref{eq:def_Pade} we can consider a second case in which the expansion converges faster than $a_{(1,1)}$, that is
\begin{equation}\label{expansionnuovaANCORA}
a_{(1,2)}(\tau)=\frac{\beta_0+\beta_1^\prime \tau}{1+\tau+\beta_3 \tau^2}\,.
\end{equation}
The picture of Eq. \eqref{expansionnuovaANCORA} is motivated by the fact that the scale factor should reproduce cosmic data at high redshift. Among all possible choices, the aforementioned one corresponds to another suitable landscape in depicting the evolution of the luminosity distance at redshift much larger than $z>1$.

In the case of \eqref{expansionnuovaANCORA}, the \emph{input} and \emph{output} limits read
\begin{subequations}
\begin{align}
a_{(1,2)}(t=-\infty)&=0\,,\\
a_{(1,2)}(t=+\infty)&=\beta_0\,,
\end{align}
\end{subequations}
as well as $da_{(1,2)}/dt|_{t=-\infty}=da_{(1,2)}/dt|_{t=+\infty}=0$.

\begin{figure}[h!]
\begin{center}
\includegraphics[width=3.2in]{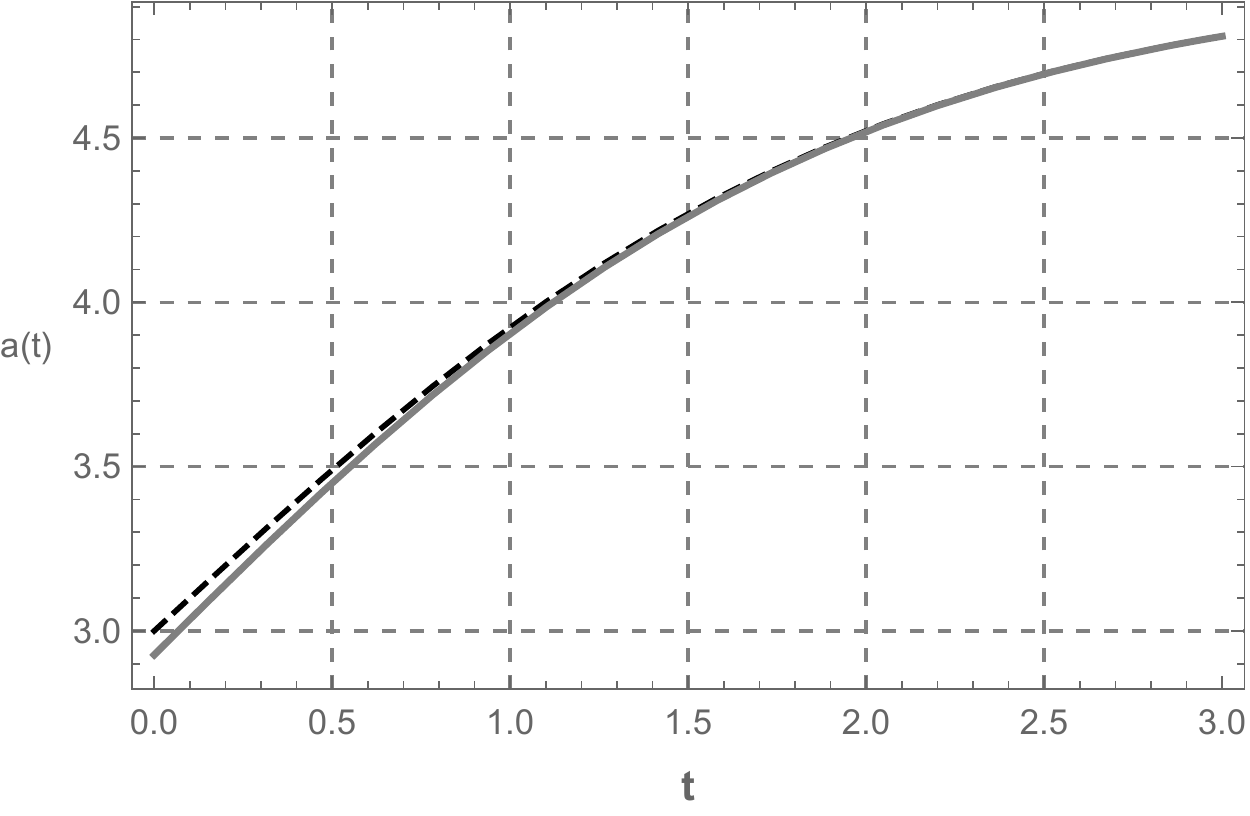}
\caption{Behavior of $a_{(1;1)}$ (dashed black curve) and $a_{(1;2)}$ (thick gray curve) as function of the cosmic time $t$. Here we adopted the indicative values $\beta_0=5$ and $\beta^\prime=1$. With the here-adopted constraint $\beta_3=0.05$, the behavior of the two scale factors is slightly different only at small times. This guarantees the Hubble rate is preserved as a genuine de-Sitter expansion rate at small redshift.}
\label{aditi}
\end{center}
\end{figure}

Unfortunately the Dirac equation \eqref{dirac} with $a_{(1,2)}(t)$ cannot be solved analytically.
Before resorting to numerics, we have to notice that
here we have an additional new free term, namely $\beta_3$. However, we require that $\beta_3$ is fixed to guarantee that the present limits over $H$ and $q$ are still valid. For being compatible with current observations and in particular to have that the present value of the Hubble parameter, i.e. $H_0$,
 lies in the domain predicted by the Planck satellite \cite{planck}, it is possible to have the tight interval: $\beta_3\in[10^{-2};10^{-1}]$. This slightly modifies the evolution of $a(t)$ in \eqref{expansionnuovaANCORA} with respect to  \eqref{expansionnuova}, as shown in Fig. \eqref{aditi}.

\begin{figure}[h!]
\begin{center}
\includegraphics[width=3.2in]{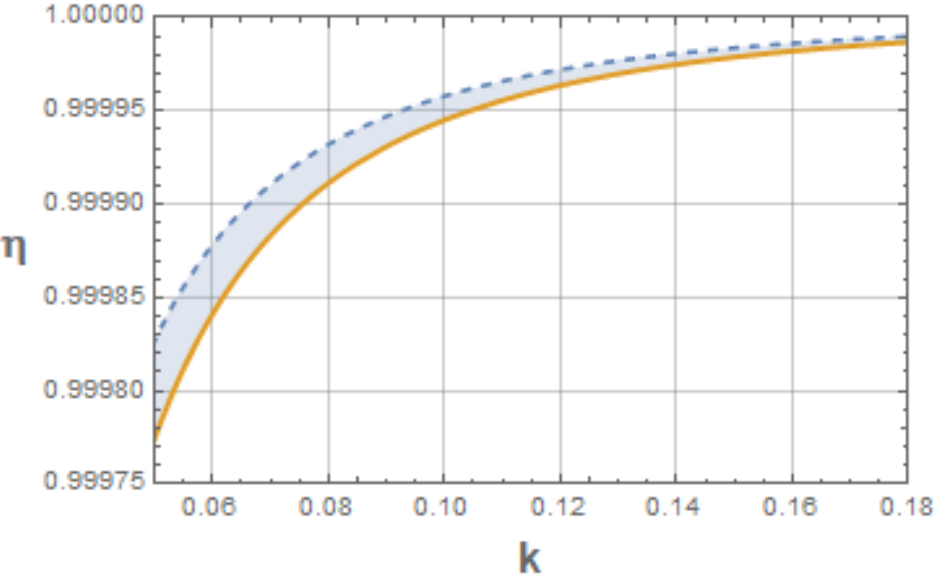}
\caption{Difference between entanglement entropy computed by the Pad\'e expansions $(1,1)$ (dashed line) and $(1,2)$ (thick line). The parameters here used are $\beta_0=5$, $\beta^\prime=1$ and $\beta_3=0.05$, with $m=10^{-2}$.}
\label{difference}
\end{center}
\end{figure}

In the case of the Pad\'e $(1,2)$ expansion, the main deviations from the $(1,1)$ case are essentially depending upon the values of $k$, although the functional forms of $S$ is unaltered. The discrepancy between entanglement entropies of the two cases has been portrayed in Fig. \eqref{difference}.
The discrepancy is extremely small even for different choices of $m$. Precisely, by shifting up the mass magnitude, discrepancies become smaller for fixed $k$.

\begin{figure}[h!]
\begin{center}
\includegraphics[width=3.2in]{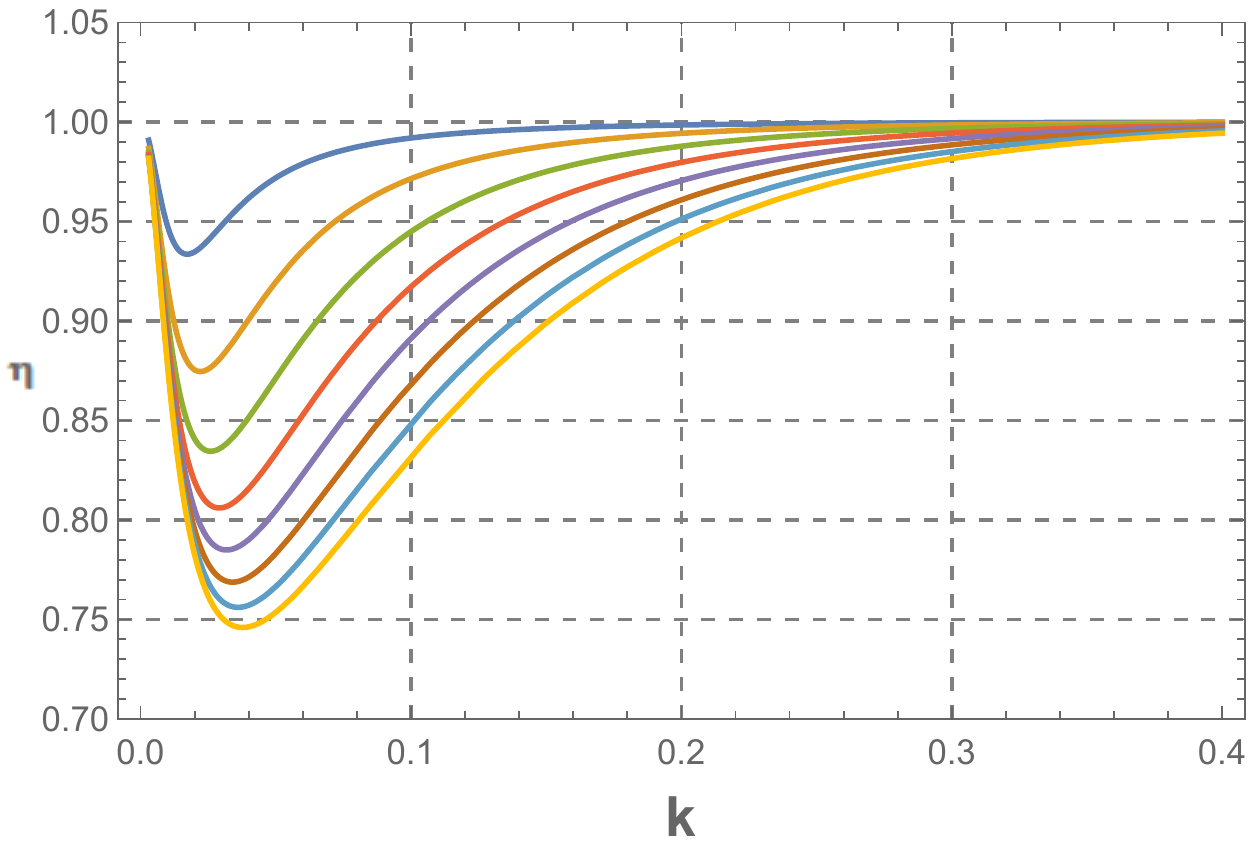}
\includegraphics[width=3.2in]{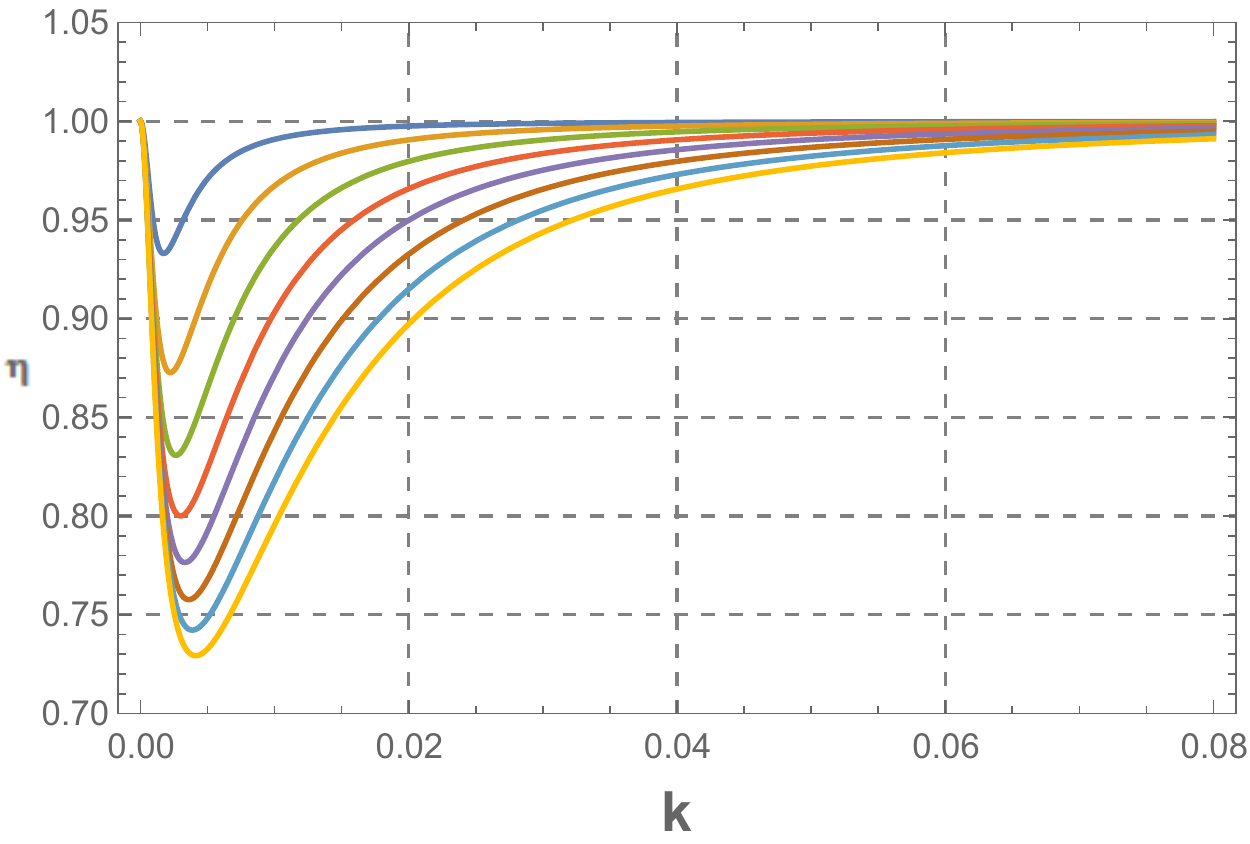}
\caption{Behavior of the transmissivity $\eta$ as function of the momentum $k$. Here we consider $\beta^\prime=1$ and $\beta_0\in[1,15]$ with step 1. The left is obtained with $m=10^{-2}$, while the right with $m=10^{-3}$.}
\label{tras1bis}
\end{center}
\end{figure}

%%%%%%%%%%%%%%%%%%%%%%%%%%%%%%%%%%%%%%%%%%%%%%%%%%%%%%%%%%%%%%%%%%%%%%%%%%%%%%%%%%%%%%%%%%%%%

\section{Final outlook and perspectives}

We dealt with entanglement production for Dirac field in a homogeneous and isotropic universe assuming a model-independent framework for the scale factor.
To do so, we considered two phases of universe's evolution corresponding to past and future respectively. We thus proposed to use the Pad\'e approximation on $a(t)$ built up in terms of \emph{functions of cosmic time} $t$, fulfilling the fewest number of basic requirements dictated by current observations over the variations of $a(t)$. We computed Pad\'e expansions (1,1) and (1,2) of the scale factor and we evaluated the asymptotic solutions of the Bogolubov transformations. Since the free constants entering our models are essentially influenced by
cosmic observations, we have thus evaluated the particle-antiparticle entanglement in the output region
according to current observations.

The results might appear qualitatively similar to those found in Ref. \cite{Fetal10}, however we here gave a physical interpretation of $a(t)$ evolution, without considering any \emph{ad hoc} toy models. Further, in our approach we are led to take a single free parameter, ($\beta_0$), to vary, while in Ref. \cite{Fetal10} one was forced to have two ($\rho$ and $\epsilon$). Going on we have considered the (1,2) order of the scale factor expansion through Pad\'e approximants. In this case we numerically determined the solution of dynamical equation to arrive at particle-antiparticle entanglement in the output region and we gave a tight range of values for $\beta_3$. Higher orders than (1,1) did not change significantly the results. Hence, we conclude that any orders higher than (1,1) do not furnish relevant deviations on $S$, if one guarantees cosmic requirements to be valid. Thus, any toy models developed in the literature are forced to be built up in terms of functions quite similar to (1,1) Pad\'e rational expansion.

The use of (1,1) order expansion is qualitatively justified for investigating quantum information processing tasks. For instance, following Ref. \cite{MPW14} we could have considered how well information is transferred from remote past to far future. It is remarkable to note that our Pad\'e expansions for small $\tau$ reduce to $a\sim \tau^p$ with $p=2;3$ respectively for the (1,1) and (1,2) expansions. This enables one to expect inflationary phases at small $\tau$, leaving unaltered the universe expansion.
The noise imparted to Dirac particles by the evolution of the universe still results equivalent to an amplitude damping channel. The transmissivity parameter  turns out to be $\eta=1-n/2$ and its behavior is reported in Fig. \eqref{tras1bis}.

At the end it is also worth mentioning that the model of Ref.\cite{BD80} can be recovered from (1,1) Pad\'e approximant, but with parameters values far away from fitting the observations.

Looking ahead it could be interesting to extend the presented approach to entanglement creation in non-homogeneous spacetimes \cite{PMM17}.
Above all it would be interesting to work directly on the universe dynamics,
considering the Wheeler-DeWitt equation. This scenario
avoids the use of Bogolubov transformations, showing
several advantages among which the possibility to constrain $\dot{a}$, without setting it to zero.

\appendix

\section{Properties of Hypergeometric functions relevant to our solutions}

Our solutions, $f_{in}^{(\pm)}(t)$ and $f_{out}^{(\pm)}(t)$, are written in terms of hypergeometric functions of the second kind, i.e.  ${}_2\mathcal{F}_1$, that can be expanded as \cite{NU88}
\begin{equation}\label{wittm}
{}_2\mathcal{F}_{1}(\theta_1^\pm,\theta_2^\pm,\theta_3^\pm,\ell(t))\approx 1+F_{1,0}\,\ell(t)+F_{2,0}\,\ell(t)^2\,,
\end{equation}
with
$F_{1,0}\equiv \frac{\theta_1^\pm\theta_2^\pm}{\theta_3^\pm}$ and $F_{2,0}\equiv \frac{\theta_1^\pm (1 + \theta_1^\pm) \theta_2^\pm (1 + \theta_2^\pm) }{2 \theta_3^\pm (1 + \theta_3^\pm)}$, when $\ell(t)\ll1$.

Another important property is the following \cite{NU88}:
\begin{align}\label{A1}
{}_2\mathcal{F}_{1}(\theta_1^{-},\theta_2^{+},\theta_3^{-},\ell_1)&=\frac{\Gamma(\theta_3^{-})\Gamma(\theta_3^{-}-\theta_1^{-}-\theta_2^{+})}{\Gamma(\theta_3^{-}-\theta_1^{-})\Gamma(\theta_3^{-}-\theta_2^{+})} \,
{}_2{\cal F}_1
(\theta_1^{-},\theta_2^{+},\theta_1^{-}+\theta_2^{+}-\theta_3^{-}+1,\ell_2)\notag\\
&+\frac{\Gamma(\theta_3^{-})\Gamma(\theta_1^{-}+\theta_2^{+}-\theta_3^{-})}{\Gamma(\theta_1^{-})\Gamma(\theta_2^{+})} \,
\ell_2^{\theta_3^{-}-\theta_1^{-}-\theta_2^{+}}\notag\\
&\hspace{1cm}\times{}_2\mathcal{F}_{1}(\theta_3^{-}-\theta_1^{-},\theta_3^{-}-\theta_2^{+},\theta_3^{-}-\theta_1^{-}-\theta_2^{+}+1,\ell_2).
\end{align}

Finally, hypergeometric functions of the second kind also satisfy the relation
\begin{equation}\label{wittw}
{}_2\mathcal{F}_{1}(\theta_1^\pm,\theta_2^\pm,\theta_3^\pm,1)=\frac{\Gamma(\theta_3^\pm)\Gamma(\theta_3^\pm-\theta_1^\pm-\theta_2^\pm)}
{\Gamma(\theta_3^\pm-\theta_1^\pm)\Gamma(\theta_3^\pm-\theta_2^\pm)}\,,
\end{equation}
known as \emph{Gauss' hypergeometric theorem} \cite{NU88}. Since Eq.\eqref{wittw} is valid for arbitrary sets of coefficients $\theta_1^\pm,\theta_2^\pm,\theta_3^\pm$, it will be also valid at asymptotic regime. Hence solutions \eqref{ameno}-\eqref{bimeno} can be in principle obtained through its use as well.

%%%%%%%%%%%%%%%%%%%%%%%%%%%%%%%%%%%%%%%%%%%%%%%%%%%%%%%%%%%%%%%%%%%%%%%%%%%%%%%%%%%%%%%

\section*{Acknowledgments}
This work has been supported by FQXi under the programme ``Physics of Observer 2016".
The work is supported in part by the Ministry of Education and Science of the Republic of Kazakhstan, for the Program 'Fundamental and applied studies in related fields of physics of terrestrial, near-earth and atmospheric processes and their practical application' IRN: BR05236494.

\end{document}